# Time-resolved reversible optical switching of the ultralow-loss phase change material $Sb_2Se_3$

Daniel Lawson[1,2], Daniel W. Hewak[2], Otto L. Muskens[1] and Ioannis Zeimpekis[2]

[1] Physics and Astronomy, Faculty of Engineering and Physical Sciences, University of Southampton, Southampton, United Kingdom
[2] Zepler Institute, Faculty of Engineering and Physical Sciences, University of Southampton, Southampton, United Kingdom

E-mail: izk@soton.ac.uk

25th of November 2021

**Abstract**

The antimony-based chalcogenide $Sb_2Se_3$ is a rapidly emerging material for photonic phase change applications owing to its ultra-low optical losses at telecommunication wavelengths in both crystalline and amorphous phases. Here, we investigate the dynamical response of these materials from nanoseconds to milliseconds under optical pumping conditions. We apply bichromatic pump-probe transient reflectance spectroscopy which is a widely used method to study the optical performance of optical phase change materials. Amorphous regions of several hundreds of nanometers in diameter are induced by pulsed excitation of the material using a wavelength of 488 nm above the absorption edge, while the transient reflectance is probed using a continuous wave 980 nm laser, well below the absorption edge of the material. We find vitrification dynamics in the nanosecond range and observe crystallization on millisecond time scales. These results show a large five-orders of magnitude difference in time scales between crystallization and vitrification dynamics in this material. The insights provided in this work are fundamental for the optimisation of the material family and its employment in photonic applications.

## 1. Introduction

The demand for switchable and adaptive components in photonics is ever increasing. [1-3] The integration of phase change materials (PCM) in photonic integrated circuits (PICs) and the likes thereof stem from their ability to display contrasting optical and electrical properties between their different crystallographic states, which include crystal and amorphous phases.[4] The reversibility of the transitions between the crystal states enables the development of components to be integrated within nanophotonic structures to achieve tuneability or switchability. The variation of intrinsic material properties brought about by stoichiometric variation and doping [5-12], which dominate the crystallisation processes, allow PCMs to be tuned to fit a wide range of applications; metasurfaces [13-16], optical memories [17-21] and reconfigurable photonic integrated circuits [22-25] to name a few. Moreover, the non-volatile nature of the PCM states means that no power is required to maintain their state. Desireable traits of PCMs in these applications include high optical contrast between states, multiple distinguisable states and high durability, the latter here referring to the retention of characteristic properties after many operation cycles.

A commonly used material for optical memories and signal routing in photonic integrated circuits is Ge-Sb-Te (GST), namely due to its large refractive index contrast between its rocksalt crystalline phase and amorphous phases, at wavelengths across both the visible and near-infrared regions of the electromagnetic spectrum (e.g $\Delta n \sim 3.5$ for $Ge_2Sb_2Te_5$ stoichiometry at $\lambda$ =1550nm). GST also displays fast switching speeds and high durabilities. However, the large extinction coefficients of GST in the near infrared and telecommuncations bands limit its integration in photonic components. Recently, antimony-based sesquichalcogenide PCMs (i.e $Sb_2Te_3$, $Sb_2Se_3$, $Sb_2S_3$) have garnered interest for photonic applications due to their vanishing extinction coefficients in the telecommunications bands and their potential application as reprogrammable components for photonic integrated circuits. [22-25]

Here, we investigate the nanosecond to millisecond time-resolved switching characteristics of a 30-nm thin film of $Sb_2Se_3$ PCM. In order to access the transient dynamics during the phase change process, we constructed a pump-probe static tester with pulsed optical writing and fast time-resolved detection using a 980 nm continuous wave probe laser. Similar techniques were introduced by Simpson and co-workers [26] to resolve the temporal dynamics of crystallization and amorphization of GST for different pump pulse powers and time duration. Recent time resolved studies revealed the enormous variation of crystallization kinetics in different PCM compositions. [27] Our studies show that the dynamical response of the $Sb_2Se_3$ PCM ranges from nanoseconds for vitrification to milliseconds for the crystallization kinetics.

## 2. Background

The antimony-based chalcogenides $Sb_2S_3$ and $Sb_2Se_3$ have received little attention until recently in phase change applications. The emergence of interest in optical PCMs with low optical losses in the visible and near-infrared spectrum has opened up new studies in these materials. The PCM $Sb_2Se_3$ combines a lower band gap energy and higher melting temperature than $Sb_2S_3$. $Sb_2Se_3$ displays comparatively low pulse energy thresholds for both set and reset operation via optically induced switching.[28] Low-loss reconfigurable MMIs were demostrated by our team employing an $Sb_2Se_3$ cladding as an active layer to control optical signal routing through a 2x2 MMI structure with low crosstalk and insertion losses.[24] In the work by Rios et al both unbalanced 2x2 MZI and micro-ring resonators used $Sb_2Se_3$ to demonstrate both multilevel phase control and optical device trimming respectively switched by microheaters.[25]

Though GST displays a much higher phase shift per unit length of PCM, when used as a cladding on silicon waveguides, the attenuative losses associated with its high extinction coefficient at the telecommuncations C-band limit its use. Due to poor scaleability (for GST-225 the phase shift induced per unit loss is two orders of magnitudes smaller than that of $Sb_2Se_3$, being 0.282 and 29.0 rad $dB^{-1}$ respectively).[28] Thus $Sb_2Se_3$, alongside $Sb_2S_3$, proves to be a suitable material where device scaleability is prioritised over footprint and optical loss is a key parameter. $Sb_2Se_3$ has also been explored as a material for structural colors enabled by switchable all-dielectric PCM metasurfaces alongside $Sb_2S_3$ and GeTe.[29] The requirements for optical PCMs are very different from their electronic counterparts and involve new definitions of performance figures in the electro-optical-thermal material response. [30]

### 2.1 Physical and optical properties of $Sb_2Se_3$

In the crystal phase, $Sb_2Se_3$ possesses an orthorhombic *Pnma* crystal structure.[31,32] The prevalence of large Peierls-like distortions in the amorphous phase make covalent bonding the most stable, and as such covalent bonds are preferentially formed as opposed to the type of 'metavalent' bonds seen in $Sb_2Te_3$.[33-35] In the amorphous phase $Sb_2Se_3$, like $Sb_2Te_3$ and $Sb_2S_3$, tends to form Peierls-distorted octahedral motifs.[31,32] Reference [31] demonstrates that the octahedral motifs around Se centres in $Sb_2Se_3$, similarly to S centres in $Sb_2S_3$, possess larger local structural deviations in the amorphous state relative to its crystalline structure. Both the tendency to form heteropolar bonds and the deviation in the short-range order between phases contribute to high thermal stability of both the crystalline and amorphous states.[31]

### 2.2 Operating principles of optical phase change materials

The intricate balance between fast crystallization and vitrification dynamics have put chalcogenides at the center of attention for their capability to produce fast non-volatile switching devices. [36,37] Amorphisation operations, or 'RESET' operations, rely on the melting of the PCM by achieving temperatures above the melting temperature $T_m$, then rapidly quenching below the glass transition temperature $T_g$. Optically induced amorphisation is achieved by using pulses of high power and short duration, typically in the order of nanoseconds. The local geometry and its physical parameters, such as heat conductivity and heat capacity, define the quenching rate which has to be faster than the crystallisation rate at temperatures above the glass transition temperature to prevent recrystalisation.

The 'SET' operation refers to the transition from an amorphous state to a crystalline state and entails heating of the PCM above its crystallisation temperature $T_c$. This is typically a slower process than amorphisation that requires less power. In applications which do not require fast modulation of the material phase a slower crystallisation speed is preferred as this reduces the requirements of the surounding materials to accommodate the subsequent amorphisation. Commonly, PCMs with fast crystallisation dynamics are deposited in thin films, typically tens of nanometres thick, on top of materials with high thermal conductivities, in order to achieve a melt-quenched amorphous phase. Because of this, the optical contrast between switched states of the PCM in films of such thickness is small due to shorter interaction lengths, which makes it hard to distinguish fully crystallised and melt-quenched states, let alone any intermediate states that are accessible by partial crystallisation.

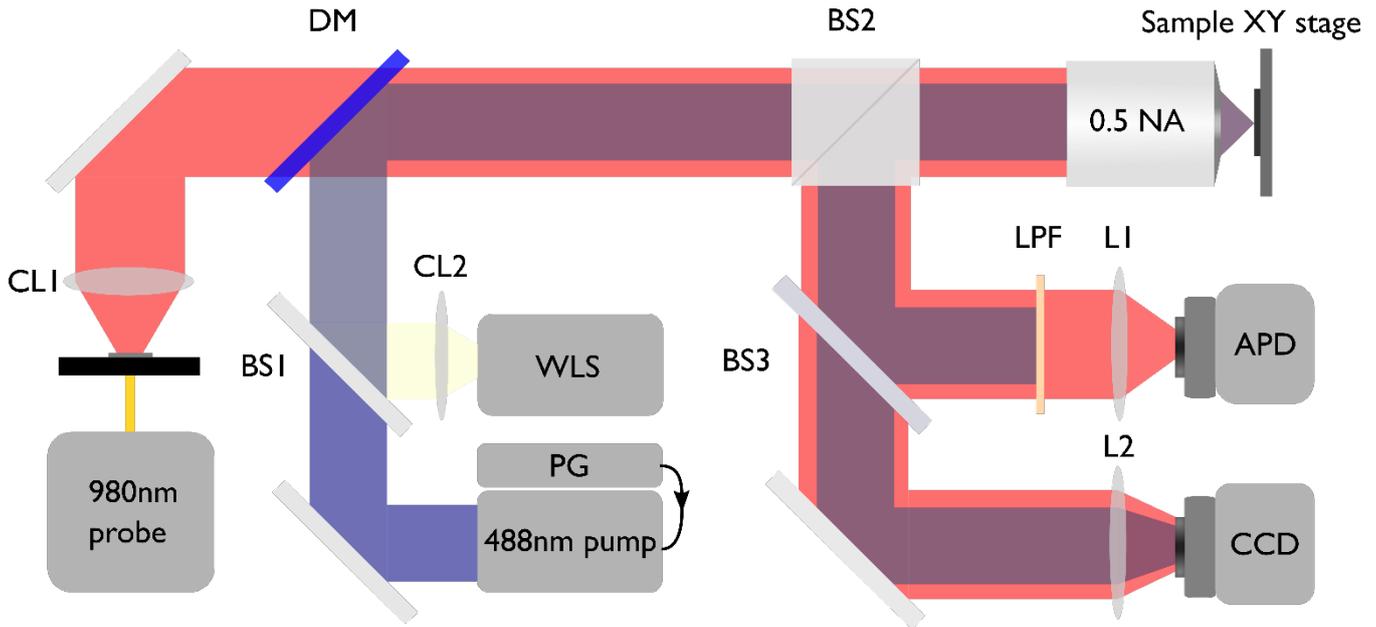

*Figure 1: Schematic of the experimental setup used for the optical switching. DM, dichroic mirror; BS, beam splitter; 0.5 NA objective lens; LPF, long pass filter; APD, avalanche photodetector; CL, collimating lens; PG, pulse generator; WLS, white light source; CCD, image sensor.*

In-situ measurement of the crystallisation dynamics, mediated by optical excitation, inherently becomes more difficult; the interpretation of optical data often has to be undertaken in conjunction with finite-element method thermal modelling in order to predict material behaviour, wherein it is hard to couple adequately the aforementioned mechanisms. More quantitative methods have been proposed, for example involving fast-heating calorimetry [38], however their implementation is considerably more complex.

## 3. Methods

### 3.1 Sample fabrication

Thin films of $Sb_2Se_3$ with thickness of 25 nm were deposited on silicon wafers with (100) orientation, and were capped with a $ZnS:SiO_2$ layer with 20%:80% composition and 200-nm thickness.[28] Both materials were deposited by means of RF magnetron sputtering (ATC-Orion 5). The capping layers were deposited using a pressure of 2 mTorr and a sputtering power of 100W without exposure to atmosphere to prevent oxidation of the PCM layer and to prevent evaporative losses of selenium during sample evaluation.

### 3.2 Bichromatic transient reflectance spectroscopy

Figure 1 shows a schematic diagram of the optical experiment configuration. The optical switching experiments were performed using a 488 nm pulsed diode, with a maximal power of 150 mW, as the source of the switching pulses, and a 980 nm diode laser operated in continuous wave as the probe laser. The power of the 980 nm probe laser incident on the samples was fixed at 15 mW to ensure that heating induced by the probe is negligible. This wavelength was chosen for the probe beam as the optical absorption coefficient of the $Sb_2Se_3$ films falls off toward the infrared, as does that of the silicon substrate. The output of the 488 nm wavelength laser was electrically modulated by a programmable pulse generator (PG) (Berkeley Nucleonics, PG in Figure 1) in order to gate optical pulses of varying power and length for both crystallisation and amorphisation. The maximum power achievable on the sample was 66 mW as measured by an optical power meter. A 100x Mitutoyo 0.55 NA objective lens (OL) was used to achieve two diffraction-limited laser spots which are spatially overlapped on the area of interest of the samples. A 500 MHz InGaAs avalanche photodetector (Thorlabs) was used to measure the reflected probe signal in combination with a 550 nm longpass filter (LPF) which is used to reject any reflected pump light. The photodiode response is relayed to a low-noise signal amplifier then to an oscilloscope (Picoscope 5244D) with 200 MHz bandwidth and 1.75 ns rise time. A white light source (WLS) is used to image the sample surface during alignment and it is switched off during transient measurements. The sample was positioned by means of a motorized XY translation stage (Thorlabs). The sample was mounted on a precision 2D tilt stage to adjust for drift of the laser focus during scans.

*3.3 Photothermal simulations*

In this work we combined photothermal and heat flow simulations of pulsed laser heating to predict the switching performance of the demonstrated materials. We employed finite element method in the multiphysics simulation software COMSOL. The optical properties of the $Sb_2Se_3$ and $ZnS:SiO_2$ layers were obtained via ellipsometry.[28] The simulation assumes that the overall structure is semi-transparent and behaves in accordance with Beer-Lambert law, such that materials with non-zero complex refractive index k attenuate an incoming laser source primarily through optical absorption. In short, the simulation considers a Gaussian beam focused on the chalcogenide layer along the z-axis. The variation of the laser intensity $I$ as a function of the depth in a material $i$ is described by the Beer-Lambert law in differential form

$$\frac{\partial I}{\partial z} = \alpha_i I$$

where $z$ is the depth of the beam in a material $i$, with temperature-independent absorption coefficient $\alpha_i$. Since the intensity changes radially and along the beam axis the spatial intensity distribution can be written as

$$I(x, y, z) = \frac{P_0}{\pi \omega(z)^2} \exp \frac{-2(x^2 + y^2)}{\omega(z)^2}$$

The heat deposited by the laser beam can then be represented by a heat source $Q$, which is the heat generated in the material as a function of local intensity $I$ and the Cartesian coordinates as written above. This results in a partial differential equation for the temperature distribution which we must also solve for

$$C_p \rho \frac{\partial T}{\partial t} = -\nabla \cdot \phi + Q$$

where $C_p$ is the specific heat capacity at constant pressure, $\rho$ is the material density and $\phi$ is the laser-induced heat flux. A Heaviside step function is used to gate the emulated optical square pulses incident on the sample in time.

## 4. Results

*4.1 Pulsed optical heating simulation of $Sb_2Se_3$ thin films*

Figure 2 shows results of a pulsed heating simulation using a pulse duration of 500 ns and varying powers between 60 mW and 90 mW. These conditions were chosen to emulate the typical process of amorphisation of a crystalline $Sb_2Se_3$ PCM layer of 25 nm thickness on top of silicon and covered with a 200 nm of $ZnS:SiO_2$ capping layer. The simulated laser beam of 488 nm wavelength was focused to a spot size of 1 μm in diameter at the interface between the silicon and the PCM. Values of 1 $Wm^{-1}K^{-1}$ and 0.15 $Jg^{-1}K^{-1}$ were used for the thermal conductivity and specific heat $C_p$ of $Sb_2Se_3$ respectively.[39,40] The simulation volume was reduced to a single quadrant by using appropriate reflecting boundary conditions. The top panels in Figure 2 show snapshots of the lateral temperature profile inside the PCM layer quadrant, where the saturated colour map illustrates the region above the melting temperature of 880 K in white colour.

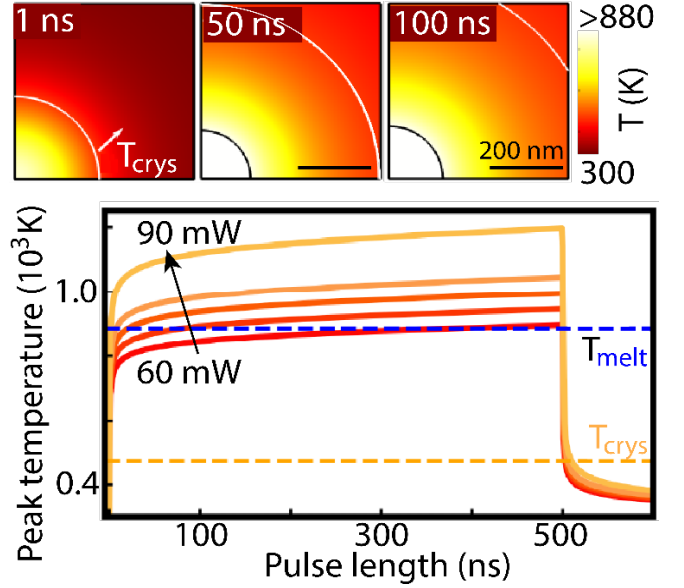

*Figure 2: Top: Snapshots of the temperature distribution at the PCM-ZnS:SiO$_2$ interface for a 500ns pulse at a pump power of 90mW immediately after the duration of the pulse. The crystallisation temperature $T_c$ (473K) is represented by the white isotherm and the melt temperature $T_m$ (884K) represented by the black isotherm. Bottom: Peak temperature at the centre of the Sb$_2$Se$_3$ thin film for 500ns pulses of varying power. The blue dashed line indicates $T_m$ and the orange dashed line represents $T_c$.*

We see that during an amorphisation pulse the peak temperature in the centre rises quickly, followed by a much slower slope related to the lateral heat flow outward from the optical pump focus. After this initial fast rise time, the region of the molten volume gradually increases in size. We can therefore see that the optical power is the main factor determining the critical threshold in the centre of the excitation volume for heating times <50 ns, with a smaller secondary contribution of the pulse duration in the hundreds of nanosecond range.

Switching off the optical heating (at 500 ns in Figure 2) results in a fast cooling of the PCM layer on a time scale of nanoseconds which is related to the out-of-plane cooling and heat loss through the surrounding layers. This fast component quenches the material to below $T_g$, indicating that the surrounding layers have sufficient capacity to take up the heat from the PCM layer and effectively cool it down to an amorphous phase. Evidence of this are in the millisecond crystallisation times measured in this communication via optical probing. The fast initial cooling is followed by a slower dynamic sets-in which is related to lateral heat flow and slower dynamics of heat in the surrounding layers.

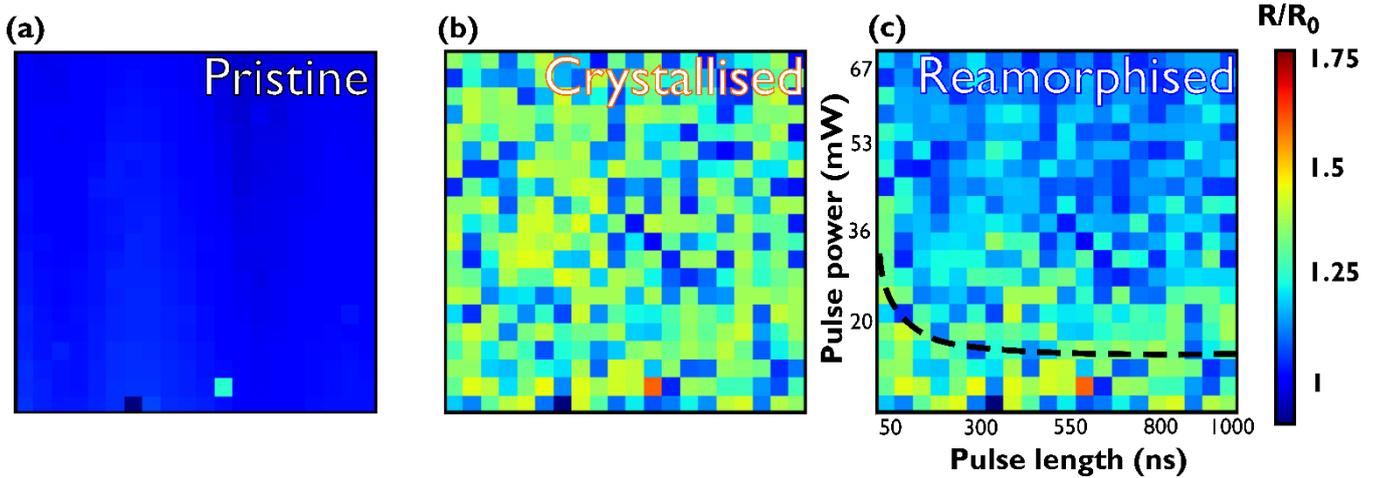

Figure 3: Measured reflectance of an array of film positions for (a) the pristine as-deposited amorphous state $R_0$, (b) the laser crystallised state $R_1$ and (c) the partial amorphous state $R_2$, normalised to the global reflectance of the as-deposited film. An array of 400 points is used to test the amorphisation thresholds in (c). The black dashed line indicates the switching threshold, above which there is a measurable reflectance change recognised as a transition between the partially amorphised final state $R_2$ and the crystalline state $R_1$.

### 4.2 Reversible switching of as-deposited amorphous $Sb_2Se_3$ thin films

In the experimental study of this work we investigate the reversible optical switching of $Sb_2Se_3$ from an as-deposited pristine amorphous phase ($R_0$) to a laser–induced crystalline state ($R_1$) and back to an amorphised state ($R_2$). All reflectance measurements presented in figures 3-5 are normalised to the average reflectance of the pristine as-deposited amorphous film ($R_0$) over the scanned area.

Figure 3 shows the results of a static tester experiment of 20×20 pulse configurations with variations of pulse length between 50 – 950 ns and pulse power from 3 – 66 mW in steps of 3 mW.

#### 4.2.1 Characterisation of the threshold amorphisation behaviour of $Sb_2Se_3$

To characterise the pulse energy required for re-amorphisation of the material we performed a parametric switching study. A 20×20 array of 1 μm$^2$ crystalline spots is first generated by using 12mW, 1s pulses spaced in a 80 μm$^2$ area of the sample by means of an XY stage. The distance of the spots guarantees adequate spacing between the crystallised areas. The reflectance change between the pristine sample and the crystallised areas was monitored and presented in Figure 3b. The amorphisation process is studied using pulses of varying length and power. The threshold switching behaviour of the amorphisation process is revealed, wherein no resolvable reflectance change is observed for pulses below a power of 15 mW, significantly lower than predicted with the pulsed heating simulation. Beyond 15 mW the induced change in reflectance tends to increase with pulse power and length, which is due to the increase of the radius of the re-amorphised mark, under the assumption that full depth vitrification is achieved in sufficiently thin films. On average, such a crystallisation event results in a change of reflectance by $ΔR_{0,1}$ ≈ 0.25. This may be attributed to local variations; selenium deficient regions may result in lower optical contrast between states. It should be noted that heat transfer to the irradiated spot is predominantly through absorption and accompanying attenuation of the laser pulses in the $Sb_2Se_3$ layer due to the high value of the extinction coefficient at 488nm ($α$ = 2.58 and 1.34 for the crystalline and amorphous phases respectively)[28]. This high dependance to the film absorption and the small size of the spot result in the local conditions of the $Sb_2Se_3$ film such as short-range coordination, number of local defects or local stoichiometry and film thickness playing a more significant role for optical switching when compared to substrate absorption.

#### 4.2.2 Time resolved switching of $Sb_2Se_3$

To observe the temporal evolution of the $Sb_2Se_3$ optical response, transient reflectance measurements were performed for both the crystallisation and amorphisation processes. Time-resolved measurements of both crystallisation of the as-deposited $Sb_2Se_3$ films and the re-amorphisation of the generated crystallised regions are shown in the first column of Figure 4. First, long optical pulses using the 488 nm pulse laser (12 mW, 1 s) are used to crystallise the film. In this case, the $Sb_2Se_3$ is heated to but not far above $T_c$. A steep increase in the reflectance is observed at the beginning of the laser pulse, which is indicative of the fast lateral crystal growth achieved initially at the centre of the irradiated area, where the film temperature is highest. As the crystal grows radially throughout the duration of the crystallisation pulse the growth rate decreases, as does the rate of increase of the reflectance. Finally, at the end of the pulse, removal of the pump excitation results in a sharp decrease of the reflectance. This is a direct result of the final cooling to ambient temperature and is attributed to a temperature dependence of the refractive index. The retention of the increased reflectance beyond the duration of the laser pulse confirms the phase-change. The growth of the crystal domains in Figure 4 can be likened to the Johnson–Mehl–Avrami–Kolmogorov (JMAK) description for an isothermal phase transformation of a finite volume, which takes the form of a sigmoid. These transients are convoluted with additional components which represent the growth of the central crystallised region at the crystal-glass interface and the induced thermo-optic effect on the refractive index.[41] The non-finite response where the two distinct epochs (i) a fast initial nucleation and growth of a central crystal domain and (ii) a slow, gradual growth of the initial are better illustrated in Figure 5(a).

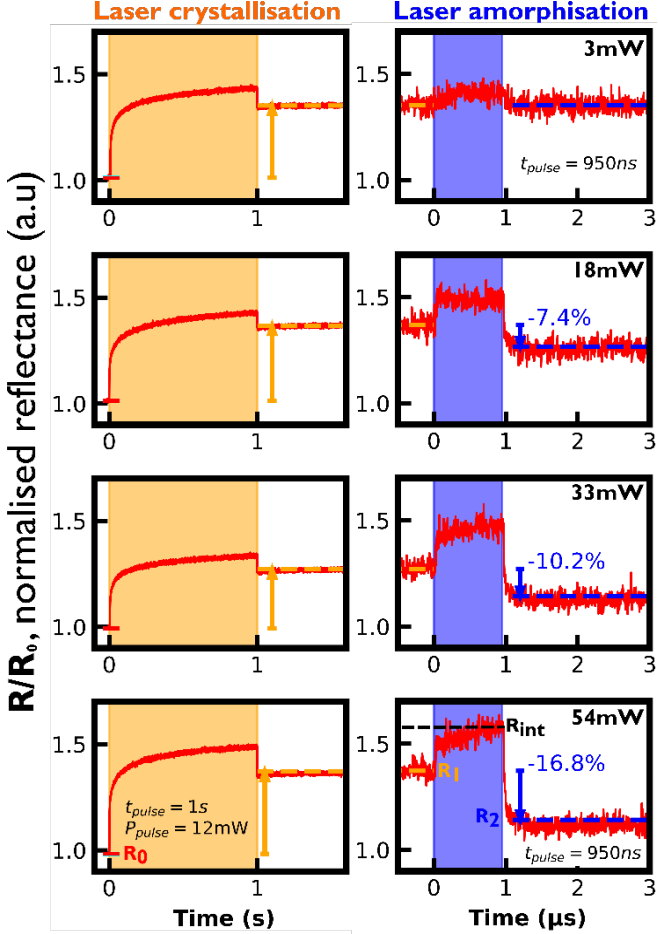

Figure 4: Time resolved reflectance measurements R/R₀ for the pristine amorphous to crystalline transition (left column) and the crystalline to partially amorphous transition (right column). The highlighted regions represent the duration of the pulse for crystallisation pulses (orange, 12mW, 1s) and amorphisation pulses (blue, various powers, 950ns). Each row represents the reversible switching of an individual pixel. Blue arrows in the right column represent the reflectance change $\Delta R_{1,2}$ expressed as a percentage value.

The right column of Figure 4 demonstrates the re-amorphisation of prior crystallised areas by 950 ns pulses of various pulse power. This pulse length is used to exemplify the behaviour commonly seen for high energy pulses. Again, at the beginning of the pulse duration the reflectivity of the sample rises due to an induced thermo-optic change in the refractive index. Removal of the optical excitation results in the quenching of the intermediate phase (highlighted in blue). At suitably high pulse powers the crystalline regions can be partially amorphised in the centre, with increasing degrees of amorphisation with increasing pulse power as in Figure 3(c). In all cases the signal remains steady after the quenching of the molten region, within a few hundreds of nanoseconds. This demonstrates the ability to achieve intermediate states by partial area crystallisation, an imperative feature for multistate memories and neuromorphic applications.

The quenching behaviours in the vitrification processes can be represented by a simple exponential decay function in Figure 5(b) for amorphisation pulses of power 60 mW and varying duration. The model that describes the evolution of the reflectance with time $R(t)$ takes the form:

$$R(t) = R_{int} \exp\left(-\frac{t}{\tau_{vitr}}\right) + R_2$$

where $R_{int}$ is the intermediate reflectance achieved at the end of the excitation, $R_2$ the reflectance of the resultant partially amorphised region and $\tau_{vitr}$ is the characteristic quenching time. Physically $\tau_{vitr}$ and hence $R(t)$, have dependence on the cooling rate of the molten region. We chose to fit this model selectively for high pulse powers as these pulses produce pronounced transient responses. $\tau_{vitr}$ is found to be around 40 ns for all pulse lengths. With the assumption that the temperatures achieved in the intermediate state are on the order of $10^3$ K, the average cooling rate ($\frac{dT}{dt}$) upon removal of the excitation is on the order of $10^8$ K s$^{-1}$. The combination of the slow crystallisation behaviour presented in Figure 5(a) and fast amorphization calculated rate reveal that $Sb_2Se_3$ can be used in a relaxed thermal constraint surrounding structure. The surrounding materials are not required to have especially high thermal conductivities to accommodate the quench.

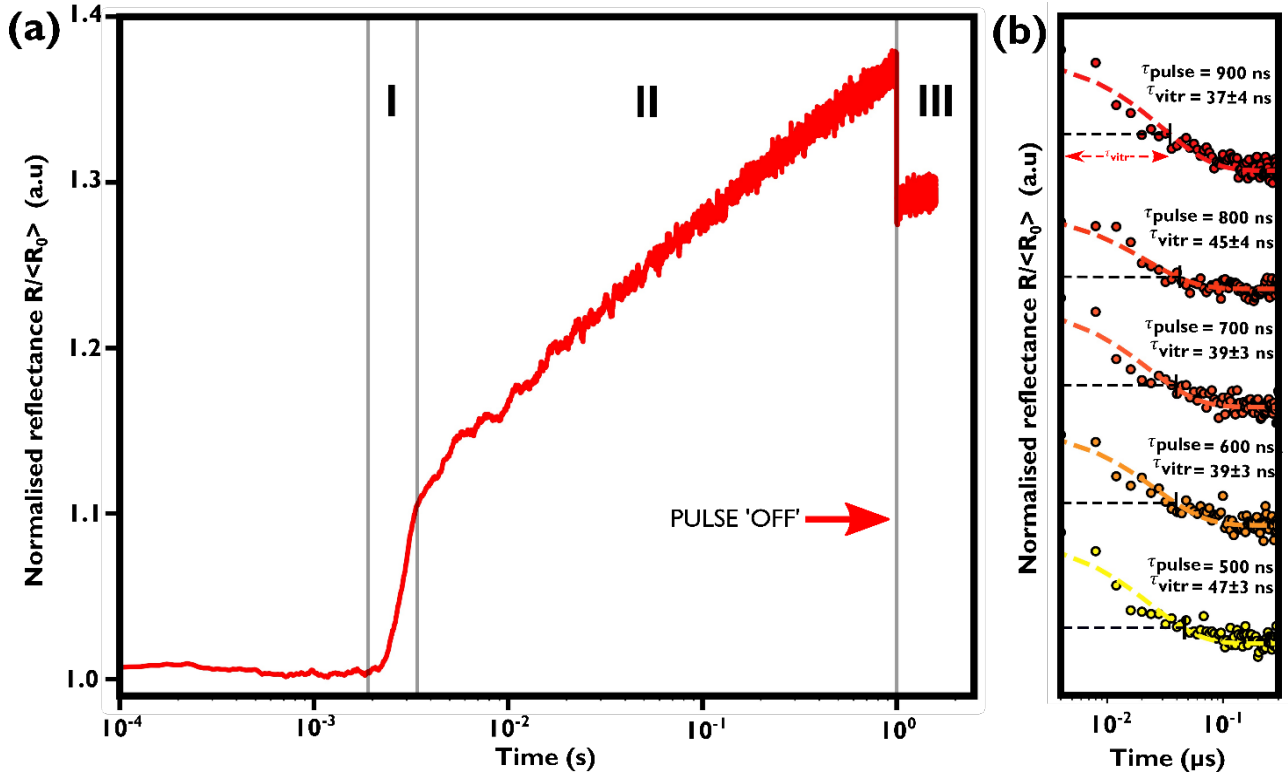

Figure 5: Logarithmic plots of time resolved reflectance measurements R/<R₀> for (a) the pristine amorphous to crystalline transition for a 12 mW 1s crystallisation pulse and (b) the crystalline to partially amorphous transition for 60mW pulses of varying length. The dashed lines in (b) represent fitted reflectance for each corresponding transient. In Figure 5 (a) the crystallisation transient demonstrates two different epochs of crystallisation in the case of these boundless amorphous films; (i) a fast initial change of the reflectance corresponding to the formation of a central crystal domain,(ii) the slow growth of the crystal domain at the crystal-glass interface and (iii) the cooling of the generated crystal and retraction of the crystal-glass interface.

## 5. Discussion

We have demonstrated the transient measurements of the phase-change of $Sb_2Se_3$ chalcogenide thin films through the use of a bichromatic reflectance spectroscopy setup. This measures the change in the refractive index of the phase change material by measuring the reflectance with a continuous laser with insufficient power to heat or modify the film. Time-resolved measurements of the reflectivity throughout the amorphisation of crystalline $Sb_2Se_3$ reveals an intermediate state induced by optical heating which increases the reflectivity of the heated region. We attribute this change in optical properties to the weakening of the van der Waals forces between covalently bonded centres during heating.

It is noteable that separate crystallisation events may exhibit small deviations from each other i.e the reflectance change achieved with the same crystallisation pulse is different. Figure 3(b) demonstrates how local film variations may influence the switching behaviour. A reflectance contrast of near 20% between the crystalline and amorphous phases is achieved purely through the use of optical pulses, with some variation in performance throughout the film. The material performance varies in the rates and spatial extent of the crystallisation for the same crystallisation pulse parameters (Figure 4). Amorphisation is achieved at much lower pulse energies than predicted, suggesting significant thermal boundary resistances at the PCM-substrate interface. Randomness in the orientation of pre-existing short-range motifs in the pristine amorphous film lead to variations in the optical absorption of the irradiated regions, which directly influence the growth rate and final size of the generated crystal domains. Small variations in the film thicknesses and variations in the defect densities may also contribute to the slight stochasticity of the crystallisation process. All of the aforementioned mechanisms have a direct impact on the film's optical properties locally, in particular the threshold energies for amorphisation and crystal growth mediated by optical pulses.

The crystallisation transient shown in Figure 5(a) demonstrates that $Sb_2Se_3$ is inherently a slow crystalliser, with crystal growth occuring around 300 ms from the initiation of the optical heating. The crystallisation process here is likened to a JMAK growth function, though with an additional component due to growth of the intial crystalline domain. In the transiemt measurements this manifests in the form of two distinct growth epochs; a fast initial crystal generation, then followed by a slow growth at the crystal-glass interface.

It is observed that the melt-quenched phase is induced at lower pulse energies than simulated, which may be attributed to the exclusion of interfacial thermal resistances between the $Sb_2Se_3$ and subsequent layers or the use of the bulk melting temperature as an indicator since the low dimensional film may display lower crystallisation and melting temperatures due to interfacial stresses. This leads to an underestimate of the PCM temperatures due to the lower thermal diffusion rates observed experimentally and the underestimation of the quenching times $\tau_{vitr}$. Experimentally we measure the full quench to be about 40 ns as opposed to the simulated quench, which occurs within a few nanoseconds.

## 6. Conclusions

Here we have demonstrated time resolved measurements of both the glass-crystal and crystal-glass transitions for $Sb_2Se_3$ through the use of a specially designed pump-probe laser microscope capable of performing transient measurements with nanosecond-scale temporal resolution. These measurements allow for the characterisation of both the crystallisation and vitrification dynamics of the PCM in response to crystallisation and amorphisation pulses respectively. We have demonstrated the reversible phase change of $Sb_2Se_3$, and show the threshold behaviour of pre-seeded crystal domains. In the crystallisation process, the onset of crystallisation in response to the optical heating occurs on the order of a few hundred milliseconds after the initation of the optical excitation, which is indicative of the slow crystallisation dynamics of $Sb_2Se_3$. The vitrification dynamics are also explored by utilising the transient optical response of the PCM under test. The quenching rate of the intermediate molten phase in the amorphisation transients, which is dependent on the sample geometry, is shown to be on the order of $\sim 10^8$ K s$^{-1}$.

This work has been primarily focussed on our understanding of the switching dynamics of phase change thin films, however the capabilities of the experimental setup can also be used to investigate the local behaviour of more complex nanoscale geometries. In future communications we intend to use this technique to explore the effect of local nanostructuring on the crystallisation and vitrification dynamics of novel PCMs.

## Acknowledgements

This work was supported by both the Engineering and Physical Sciences Research Council (EPSRC) in part through grant EP/M015130/1, Manufacturing and Application of Next Generation Chalcogenides.